\documentclass[prc,twocolumn,showpacs,showkeys,nofootinbib,superscriptaddress]{revtex4}
\usepackage{graphicx}
\usepackage{dcolumn}
\usepackage{epsfig}
\usepackage{bm}

\begin{document}

\topmargin -0.50in

\title{$0\nu\beta\beta$ nuclear matrix elements and the occupancy of individual orbits \\
}

\author{Fedor \v Simkovic},
\email{fedor.simkovic@fmph.uniba.sk}
\altaffiliation{
On  leave of absence from Department of Nuclear
Physics, Comenius University, Mlynsk\'a dolina F1, SK--842 15
Bratislava, Slovakia}
\author{Amand Faessler}
\email{amand.faessler@uni-tuebingen.de}
\affiliation{Institute f\"{u}r Theoretische Physik der Universit\"{a}t
T\"{u}bingen, D-72076 T\"{u}bingen, Germany}
\author{Petr Vogel}
\email{pxv@caltech.edu}
\affiliation{Kellogg Radiation Laboratory and Physics Department, Caltech,
Pasadena, California, 91125, USA}

\begin{abstract}
The measured occupancies of valence orbits in $^{76}$Ge and $^{76}$Se are used
as a guideline for modification of the effective mean field 
energies that results in better description
of these quantities. With them, in combination with the selfconsitent
renormalized quasiparticle random phase approximation (SRQRPA) method that ensures 
conservation of the mean particle number in the correlated ground state, we show  that the 
resulting $0\nu\beta\beta$ nuclear matrix element for the $^{76}$Ge $\rightarrow$ $^{76}$Se
transition is reduced by $\sim$25\% compared to the previous QRPA value, and
therefore the difference
between the present approach and the interacting shell model
predictions becomes correspondingly
smaller. Analogous modification of the mean field energies for the $A=82$ system
also results in a reduction of  $0\nu\beta\beta$ matrix element 
for the $^{82}$Se $\rightarrow$ $^{82}$Kr transition, making it also closer to the shell model prediction.
\end{abstract}



\date{\today}

\maketitle

\section{Introduction}

The fundamental importance of the search for  $0\nu\beta\beta$ decay is widely accepted
(see, e.g. the {\it APS Study of Physics of Neutrinos} \cite{APS}). Observing the decay would tell
us that the total lepton number is not a conserved quantity, and that, consequently, neutrinos
are massive Majorana fermions.  Experimental searches for the $0\nu\beta\beta$
decay, of ever increasing sensitivity, are being pursued worldwide (for a recent review
of the field, see \cite{AEE}). However, interpreting existing results as a measurement
of the neutrino effective mass, and planning new experiments,
depends crucially on the knowledge of the corresponding nuclear matrix elements that
govern the decay rate.  
Accurate determination of the nuclear matrix elements, and a realistic estimate of their
uncertainty, is therefore an integral part of the study.

The nuclear matrix elements for  $0\nu\beta\beta$ decay must be evaluated using
tools of nuclear structure theory. Unfortunately, there are no observables that could 
be directly linked to the magnitude of  $0\nu\beta\beta$ nuclear matrix elements
and that could be used to determine them in an essentially model independent way.
In the past, knowledge of the $2\nu\beta\beta$-decay rate and of the ordinary 
$\beta$ decay $ft$ values were used to constrain the nuclear model parameters,
in particular when the Quasiparticle Random Phase Approximation (QRPA) was
employed  \cite{us,SC}. Clearly, when other relevant data become available, and the
nuclear model is constrained to reproduce them, confidence in the deduced
 $0\nu\beta\beta$ nuclear matrix elements would increase.
Recently a set of such data,
the occupation numbers of neutron valence orbits in
the initial $^{76}$Ge and final $^{76}$Se nuclei, were determined
 in a series of measurements of cross sections for neutron adding and 
 removing transfer reactions \cite{Schiffer1}. A similar series of measurements
 involving proton transfer reactions also became recently available \cite{Schiffer2}. 
 
 Here we examine in detail how sensitive the matrix elements are to 
 these quantities, and how much the previously determined nuclear 
 matrix elements change when 
 the input of the nuclear model is modified so that occupancies of individual
 orbits are correctly reproduced.
As in the previous calculations \cite{us}, we use the QRPA method and its generalizations.

The occupation numbers for orbits with angular momentum $j$ (and any other
quantum numbers)  in the initial nucleus, measured experimentally, are simply
\begin{equation}
n^{exp}_j = \langle 0^+_{init} | \Sigma_m c^+_{j,m} c_{j,m} | 0^+_{init} \rangle ~,
\label{eq:true}
\end{equation}
and the same quantity is determined for the ground state $| 0^+_{fin} \rangle$ of the final nucleus.
Here $c^+_{j,m}$ is the creation operator for a proton in the orbit $j_p$ or a neutron
in the orbit $j_n$ and $c_{j,m}$ is the
corresponding annihilation operator.
The states $| 0^+_{init} \rangle$ and $| 0^+_{fin} \rangle$ are the true ground states
with all correlations in them.

Theoretically, for pure pairing BCS wave functions, the occupation numbers
\begin{equation}
n^{BCS}_j = \langle 0^+_{BCS} | \Sigma_m c^+_{j,m} c_{j,m} | 0^+_{BCS} \rangle = v_j^2 
\times (2j + 1) ~,
\label{eq:BCS}
\end{equation}
depend only on the amplitudes $v_{j_p}$ or $v_{j_n}$ that
are obtained by solving the gap equations. 
The amplitudes $v$ are constrained by the requirement that the expectation value
of the total neutron and proton  numbers are conserved, i.e.,
\begin{equation}
N ({\rm or} Z) = \Sigma_{n(p)} n^{BCS}_{n(p)}  \equiv \Sigma_{n(p)} v^2_{n(p)} \times (2j_{n(p)} + 1) ~.
\label{eq:BCS2}
\end{equation}
Here, and in the following, we use $n$ and $p$ to label all quantum numbers
of the corresponding neutron or proton orbits.

However, in the correlated QRPA ground state the occupation numbers are no longer 
the pure BCS quantities. Instead, they depend, in addition, on the solutions of the
QRPA equations of motion for all multipoles $J$, and can be evaluated using
\begin{eqnarray}
{\rm n}^{QRPA}_ {n(p)}& = & \langle 0^+_{QRPA} |    \Sigma_m c^+_{n(p),m} c_{n(p),m} | 0^+_{QRPA} \rangle
\\ \nonumber
& \simeq & (2j_{n(p)}+ 1) \times [ v^2_{n(p)}+ (u^2_{n(p)} - v^2_{n(p)})~ \xi_{n(p)} ],
\end{eqnarray}
where 
\begin{equation}
\xi_{n(p)}  
 =  (2 j_{n(p)} +1)^{-1/2}
\langle 0^+_{QRPA} | \left[a^+_{n(p)} a_{n(p)}\right]_{00} | 0^+_{QRPA} \rangle
\end{equation}
is the expectation value of the number of quasiparticles in the orbit $n(p)$. (Here $a^+_{j_{n(p)},m}, 
a_{j_{n(p)},m}$
are the creation and annihilitation operators for the quasiparticle with quantum numbers $n(p),m$.) 

The quasiparticle occupation numbers $\xi_{n(p)}$ can be obtained iteratively using the equations
of motion of the Renormalized Quasiparticle Random Phase Approximation (RQRPA)
and of the Selfconsistent RQRPA(SRQRPA) through the
renormalization factors ${\cal D}_{pn}$:
\begin{eqnarray}
\label{eq:QRPA}
{\cal D}_{pn}
 &  = & 1 -  \xi_p - \xi_n  \\  \nonumber
  &= & 1 -  \frac{1}{2j_p + 1} 
\Sigma_{n'} {\cal D}_{pn'} 
\left( \Sigma_{J,k} (2J+1) | \overline{Y}_{pn'}^{J,k} |^2 \right)\nonumber
\\  \nonumber
&&~~ -  \frac{1}{2j_n + 1} \Sigma_{p'} {\cal D}_{p'n} 
\left( \Sigma_{J,k} (2J+1) | \overline{Y}_{p'n}^{J,k} |^2 \right),
\end{eqnarray}
where ${\overline{Y}}^{J,m}_{pn}  =  {\cal D}^{1/2}_{pn}~ Y^{J,m}_{pn}$.

Note that the occupation numbers $n^{QRPA}_{n(p)}$ are no longer constrained by the same
requirement, Eq.(\ref{eq:BCS2}), that the particle number is conserved on average,
as the BCS occupation numbers are. 

The past applications of QRPA to the evaluation of the $0\nu\beta\beta$ nuclear matrix elements
used standard parametrizations of the nuclear mean field, usually in the form of the 
Coulomb corrected Woods-Saxon
potential fitted globally to a variety of nuclear properties (typically used parameters
are those quoted in \cite{BM} or in \cite{Bertsch}).
In few papers \cite{SC2} attempts were made to modify the single particle energy input in order to
better describe the energy levels of the $\beta\beta$-decay candidate nuclei. 
It turns out that, at least in the $A = 76$ case,  the quantities $n_{n(p)}^{BCS}$
based on the Woods-Saxon potential single
particle energies,  do not agree well with the
experimental results of \cite{Schiffer1}. In particular, the occupancy of  the 
neutron $g_{9/2}$ orbit appears
to be underestimated.

Shortly after the data of Ref.\cite{Schiffer1} became available, a new publication
\cite{SC3} appeared,
where the neutron single particle energies (only the valence orbits $g_{9/2}, f_{5/2}, p_{1/2}
{\rm ~and~} p_{3/2}$) were modified so that the neutron number occupancies 
of $^{76}$Ge and $^{76}$Se, for which the quantities $n^{BCS}_{n(p)}$ were used, 
were better reproduced. The proton mean field energies were also modified so that the
one-quasiparticle energies in the odd-Z nuclei $^{77}$As and $^{77}$Br were 
also better described. The authors of Ref. \cite{SC3} conclude that this modifications
result in  sizable reduction of the $0\nu\beta\beta$ nuclear matrix element
for the $^{76}$Ge $\rightarrow$ $^{76}$Se transition.

In the present work we carefully analyze the r\^{o}le of the constraints represented by
the knowledge of the orbit occupancies $n^{exp}_{n(p)}$.  Clearly, these quantities 
reflect presence of correlations beyond pairing correlations described by $n^{BCS}_{n(p)}$.
As a closest substitute for these correlations we use here $n^{QRPA}_{n(p)}$. Moreover,
to describe such correlations, we use the SRQRPA that, as the simple BCS,
and unlike the QRPA or RQRPA, conserves
the neutron and proton particle numbers on average. (We describe the method in more
detail in the next section).

Let us stress that the comparison between the measured and calculated occupancies
in the experimental Refs. \cite{Schiffer1,Schiffer2}, as well as in the theoretical paper \cite{SC3}, was
based on equating the experimental values $n_{n(p)}^{exp}$ with the BCS values $n_{n(p)}^{BCS}$.
As pointed out above this is not really a justified comparison as far as the QRPA and its
generalizations are concerned.
  
The paper is organized as follows. In the next section we briefly describe the SRQRPA
method, and illustrate the effect of QRPA correlations on the occupation numbers.
In Section III  we discuss the modifications of the mean field energies for the
$A$ = 76 system that
results in better description of the occupation numbers of valence orbits.
We also show that using the modified mean field energies improves the description
of the contribution of low-lying states to the $2\nu\beta\beta$ matrix element.
In Section IV we present our result for the  $^{76}$Ge $\rightarrow$ $^{76}$Se
$0\nu\beta\beta$ nuclear matrix element and show that the mean field adjustment
needed in order to better describe the orbit occupancies leads to reduction of
the difference between the QRPA and nuclear shell model results \cite{Poves}.
Analogous modifications are applied in Section V to 
the $A=82$ system with two more protons
and four more neutrons then in $A=76$. It also results in a noticeable reduction of $M^{0\nu}$
for the $^{82}$Se   $0\nu\beta\beta$ decay. We describe
the contribution of individual orbits to the $0\nu\beta\beta$ matrix element in Section VI 
and conclude in Section VII.

\section{Selfconsistent quasiparticle random phase method}

The standard QRPA method consists of two steps. First, the like-particle pairing interaction is taken
into account by employing the quasiparticle representation. In the second step the linearized
equations of motion are solved in order to describe small amplitude vibrational-like modes
around that minimum. In the renormalized version of QRPA the violation of the Pauli
exclusion principle is partially corrected. 

The drawback of QRPA and RQRPA is the fact that, unlike in BCS, and
as mentioned already earlier, the
particle number is not conserved automatically, even on average, i.e., 
\begin{equation}
\Sigma_{j_n} n_{j_n}^{QRPA} \ne N
\label{eq:ineq}
\end{equation}
and the same is true for the proton states. Naturally, in the limit of negligibly small amplitudes,
when the quasiparticle occupation numbers $\xi_p$ and $\xi_n$ in Eq.(\ref{eq:QRPA})
are small, the inequality  in (\ref{eq:ineq}) is correspondingly small as well. However,
for realistic hamiltonians the differences between the left-hand and right-hand sides
of Eq.(\ref{eq:ineq}) in QRPA is of the order of unity (an extra or missing neutron or proton).

The selfconsitent renormalized QRPA method (SRQRPA) removes this drawback 
by treating the BCS and QRPA vacua simultaneously. For the neutron-proton systems,
of interest in the present context, the method was proposed and tested on the
exactly solvable simplified models in Refs \cite{Delion}. It is a generalization of the
procedure proposed earlier in \cite{Schuck}.

Here we briefly describe the basic features of SRQRPA.  In QRPA, RQRPA
and SRQRPA the
phonon operators are defined as
\begin{equation}
Q^{\dagger (k)}_{J,M} = \Sigma_{pn} [ X_{(pn) J}^k A^{\dagger}_{(pn) J,M} -
Y_{(pn) J}^k \tilde{A}_{(pn) J,M} ] ~,
\label{eq:phonon}
\end{equation}
where $X_{(pn) J}^k$ and $Y_{(pn) J}^k$ are the usual variational amplitudes,
and $A^{\dagger}_{(pn) J,M}$ is the angular momentum coupled two-quasiparticle 
creation operator. The $X$ and $Y$ amplitudes, as well as  the corresponding
energy eigenvalues $\omega_k$ are determined by solving the QRPA eigenvalue equations
for each $J^{\pi}$
\begin{equation}
\left( \begin{array}{cc}
{\cal A} & {\cal B} \\ {\cal -B} & {\cal -A} 
\end{array} \right)
\left( \begin{array}{c}
X \\ Y \end{array} \right)
= \omega \left(
 \begin{array}{c}
X \\ Y \end{array} \right) ~.
\label{eq_rpa}
\end{equation}
The matrices ${\cal A}$ and ${\cal B}$ above are determined by the hamiltonian rewritten in terms of
the coupled quasiparticle operators:
\begin{eqnarray}
& &{\cal A}^J_{pn,p'n'} = \\  \nonumber
& & \langle 0^+_{QRPA} | [  \bar{A}_{(pn) J,M},[ \hat{H},\bar{A}^{\dagger}_{(p'n') J,M}
]] |  0^+_{QRPA} \rangle \\
\nonumber
& & {\cal B}^J_{pn,p'n'} = \\ \nonumber
& & \langle 0^+_{QRPA} | [ \bar{A}^{\dagger}_{(pn) J,-M}(-1)^M ,[ \hat{H},\bar{A}^{\dagger}_{(p'n') J,M}
]] |  0^+_{QRPA} \rangle ~,
\label{eq_ABdef}
\end{eqnarray}
where $\bar{A}_{(pn) J,M} = D_{pn}^{-1/2} A_{(pn) J,M}$.
The resulting matrices are independent of the angular momentum projection $M$.

In RQRPA and SRQRPA the nonvanishing values of
$ {\cal D}_{pn} -1$ is taken into account by using the amplitudes
\begin{equation}
{\overline{X}}^m_{(pn, J^\pi)}  =  {\cal D}^{1/2}_{pn}~ 
X^m_{(pn, J^\pi)}, ~~~~~
{\overline{Y}}^m_{(pn, J^\pi)}  =  {\cal D}^{1/2}_{pn}~ 
Y^m_{(pn, J^\pi)} ~,
\end{equation}
instead of the standard $X$ and $Y$,
everywhere and also in the QRPA equations of motion.

By doing all of this an inconsistency appears between the BCS, with the ground state 
$| 0^+_{BCS} \rangle$,
and QRPA (as well as RQRPA) with the ground state $| 0^+_{QRPA} \rangle$. In SRQRPA
this inconsistency is overcome by reformulating the BCS equations. This is achieved
by recalculating the $u$ and $v$ amplitudes.
In SRQRPA the state around which the vibrational modes
occur is no longer the quasiparticle vacuum, but instead the
Bogoliubov transformation is chosen is such a way that provides the optimal and consistent
basis while preserving the form of the phonon operator, Eq. (\ref{eq:phonon}).
The modified coefficients of the Bogoliubov transformation still fulfill the
basic requirement that the so-called dangerous graphs, terms in the
Hamiltonian with only two quasiparticle creation or annihilation operators, vanish.

In practice, the SRQRPA equations are solved iteratively. One begins with the standard BCS 
$u,v$ amplitudes, solves the RQRPA equations of motion and calculates the factors $D_{pn}$.
The $u,v$ amplitudes are recalculated and the procedure is repeated until the selfconsistency
is achieved.

The SRQRPA was applied initially to the evaluation of $2\nu\beta\beta$ matrix elements in
Ref.\cite{Bobyk1} and to the evaluation of $0\nu\beta\beta$ matrix elements in Ref.\cite{Bobyk2}.
Numerically, the double iteration procedure represents a challenging problem. 
To simplify it, in Refs.\cite{Bobyk1,Bobyk2} the bare interaction was used, and no attempt
was made to fit the odd-even mass differences. 
In addition, no adjustment of the particle-particle coupling constant $g_{pp}$ was made,
and $g_{pp} = 1$ was used. Consequently, the numerical values
disagreed noticeably with the experiment in the   $2\nu\beta\beta$ case, and with
calculations by other authors in the $0\nu\beta\beta$ case. 

The numerical problems were resolved in Ref.\cite{Benes} where instead of the G-matrix 
based interaction the pairing part (and only that part)
of the problem was replaced by a 
pairing interaction that uses a constant matrix element
whose value was adjusted to reproduce the experimental odd-even mass differences.
This is the procedure that we adopt also here, after showing that within the QRPA the
replacement of the G-matrix by a constant pairing matrix element makes little difference
(see below). Thus, in the iterative procedure only the chemical potentials 
$\lambda_n$ and $\lambda_p$ are changed. 

Adopting this simplification, and using the usual requirement, as in \cite{us},
namely that the 
$2\nu\beta\beta$ decay rate is correctly reproduced
by renormalizing the coupling constant $g_{pp}$ correspondingly, 
the authors of Ref. \cite{Benes}
have shown that the $0\nu\beta\beta$ matrix elements evaluated with SRQRPA
agree quite well with the matrix elements of Refs.\cite{us}. The requirement of conserving
the particle number have not caused substantial changes in the value of the 
$0\nu\beta\beta$ matrix elements in that case.

In Table \ref{tab:ws} we illustrate the problem of the particle number nonconservation
within QRPA and to some extent also in RQRPA, 
and its restoration in SRQRPA. The case of $^{76}$Ge $\rightarrow$ $^{76}$Se
is chosen, with $^{40}$Ca as a core, and with the $p,f$ and $s,d,g$ shells (9 single particle
orbits) for both neutrons and protons included. Note that in Ref.\cite{SC3}, as noted above, 
only the BCS 
occupancies were considered. We believe, contrary to the arguments there, that the
effect of average particle number nonconservation in the QRPA vacuum need to be
considered. Using SRQRPA, which does conserve the average particle number,
is certainly more consistent  when the problem of individual orbit occupancies
is addressed. 

Even though the conservation of the average particle number is almost restored in RQRPA
and, as we will show below, the numerical values of the $M^{0\nu}$ are very similar in
RQRPA and SRQRPA, we still prefer to use the selfconsistent method. Among other things,
the violation of the Ikeda sum rule, which in the RQRPA is as large as 20\%, is reduced
substantially (but not eliminated completely) when SRQRPA is employed.

\begin{table}[htb]
\begin{center}
\caption{ The expectation values of the particle number operator. The mean field energies as in
Ref. \cite{us}. For QRPA, RQRPA and SRQRPA the particle-particle interaction renormalization 
constant $g_{pp}$ is chosen from the requirement that the $2\nu\beta\beta$-decay rate is
correctly reproduced. }
\label{tab:ws}

\vspace{0.1cm}

\begin{tabular}{|lcccc|}
\hline
System: & BCS & QRPA & RQRPA & SRQRPA \\
\hline
initial protons & 12.00 & 13.05 & 12.05 & 12.00 \\
final protons & 14.00 & 14.61 &  14.01 &14.00 \\
initial neutrons & 24.00 & 23.86 & 23.98 & 24.00 \\
final neutrons & 22.00 & 22.16 &  21.95 & 22.00 \\
\hline
\end{tabular}
\end{center}
\end{table}

Finally, to see the difference in treating the pairing part of the problem using the realistic 
G-matrix based interaction (but adjusting its strength so that the experimental pairing gaps
are correctly reproduced, as was done in Refs. \cite{us}) and calculation performed with the
schematic pairing force (a constant matrix element adjusted similarly) we quote the QRPA
and RQRPA values of the $M^{0\nu}$ matrix element, again for the 
$^{76}$Ge $\rightarrow$ $^{76}$Se case. 
With realistic pairing interaction we obtain $M^{0\nu}$ = 4.3(3.8) with QRPA(RQRPA)
while with the schematic pairing interaction the result is 4.4(3.9). 
Using the schematic pairing makes little
difference in this case.

\section{The r\^{o}le of orbit occupancies}

The occupancies of the neutron and proton valence orbits in $^{76}$Ge and $^{76}$Se
were determined experimentally
in Refs. \cite{Schiffer1,Schiffer2}.
As shown in Section I, within QRPA and its generalizations, the occupancies of individual orbits,
corresponding to the summed spectroscopic strength measured in Refs. \cite{Schiffer1,Schiffer2},
are determined not only by the BCS amplitudes $u$ and $v$ 
but also by the quasiparticle content of the
correlated ground state $| 0^+_{QRPA} \rangle$. In turn, the BCS amplitudes $u$ and $v$ 
depend on the nuclear mean field energies, on pairing gaps $\Delta$ that are fitted to agree 
with the known odd-even mass differences, on the chemical potentials $\lambda$ that are
determined by the requirement that the particle number is conserved on average
and, within SRQRPA that we adopt, indirectly 
on the solutions of the phonon equations of motion.

In Refs. \cite{us} the mean field was based on Coulomb corrected Woods-Saxon
potential using the globally fitted parameters quoted in \cite{BM}. The resulting valence
orbits occupancies do not agree very well with experiment in that case.
In order to describe the experimental occupancies better, we modify the input mean field
energies to some extent, mainly for the valence orbits. Since our primary goal is to
evaluate the nuclear matrix elements for the $0\nu\beta\beta$ decay which depend on the
quasiparticle energies only weakly, modifying the mean field energies essentially means
that, through the $u$ and $v$, the occupancies are adjusted.

On the other hand, we use the known rate of the  $2\nu\beta\beta$-decay to fix the renormalization
constant $g_{pp}$, the strength of the neutron-proton particle-particle force. The matrix element
for the $2\nu\beta\beta$ decay, in turn, depends on the energies of the $1^+$ states significantly.
In the intermediate nucleus $^{76}$As the energies of a few low-lying $1^+$ states are known.
In calculation that used the global Woods-Saxon potential \cite{us} 
these energies were not described well 
either, indicating again that the global single-particle potential is not optimal for the $A=76$ system.

\begin{figure}[htb]
\includegraphics[width=.48\textwidth]{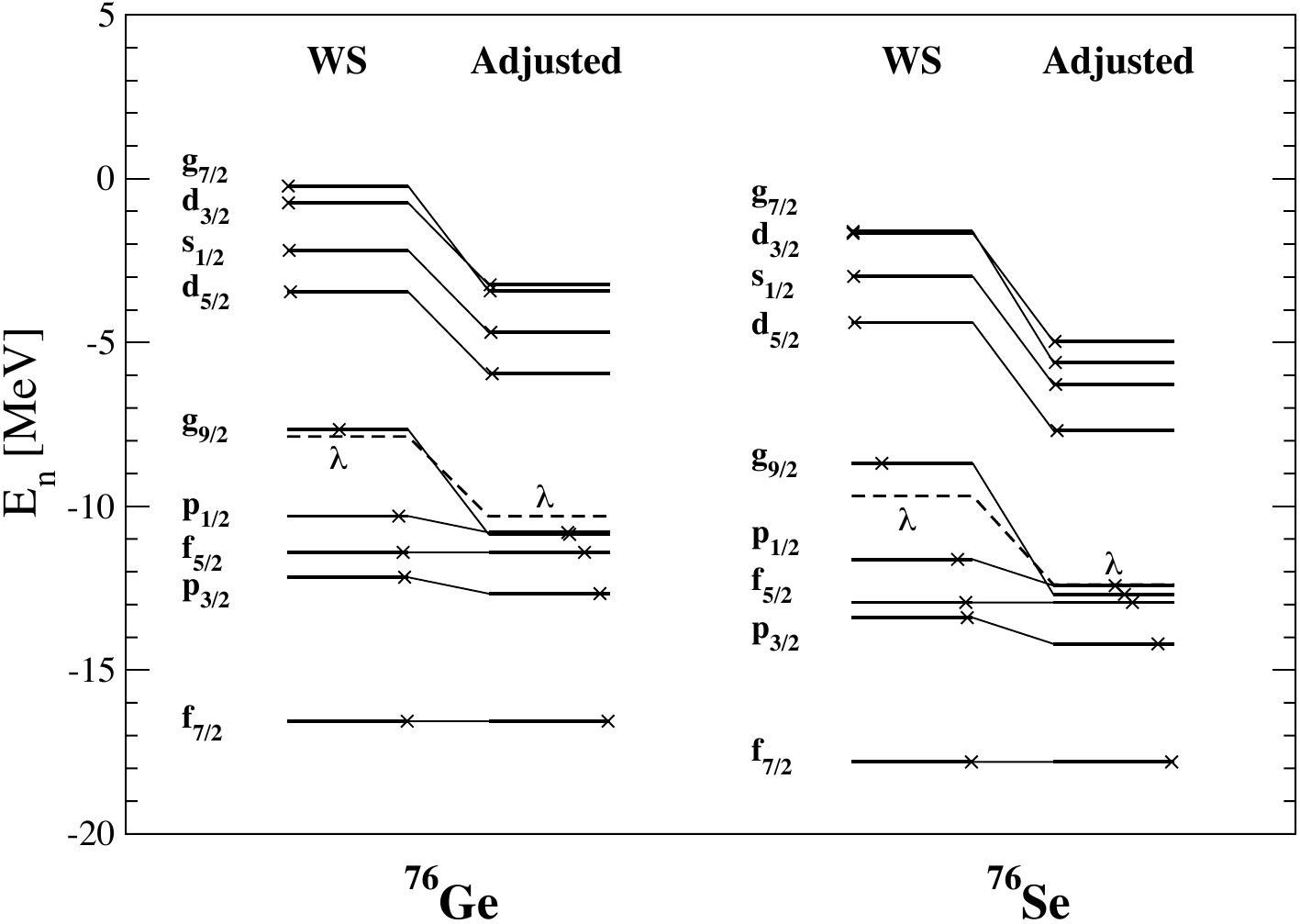}
\caption{Comparison of the neutron levels in the Woods-Saxon potential used in Refs. \cite{us} 
(WS) and the adjusted mean field energies used here
and described in the text. Symbols $\lambda$ indicate the chemical potential
and the crosses indicate the occupancy of the individual orbits.}
\label{fig:neut}
\end{figure}

\begin{figure}[htb]
\includegraphics[width=.48\textwidth]{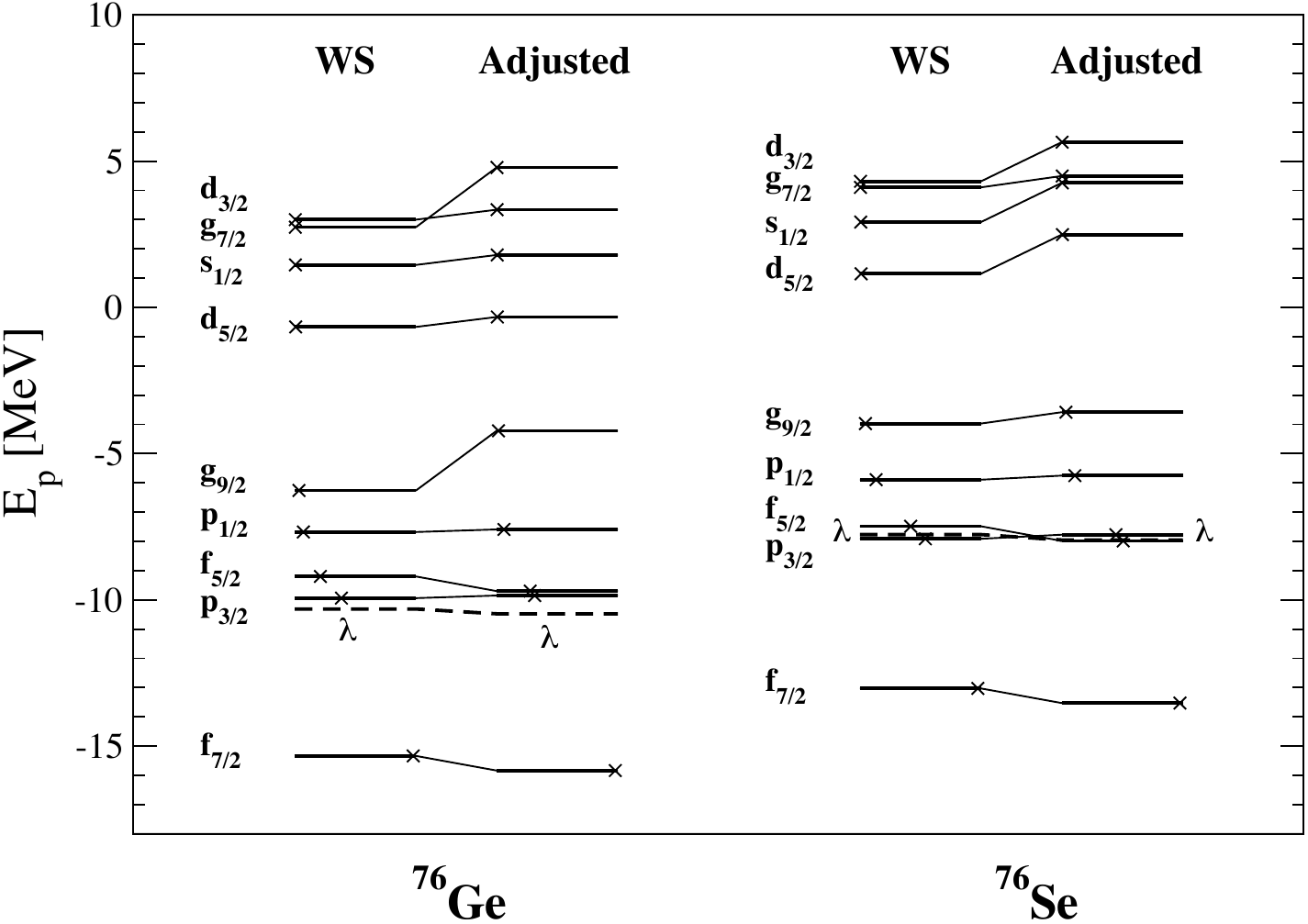}
\caption{ Comparison of the proton levels of the Woods-Saxon potential used in \cite{us}
and the adjusted mean field energies used here
and described in the text. Notation as in Fig. \ref{fig:neut}.}
\label{fig:prot}
\end{figure}

Guided by such considerations we modified the mean field energy input, determining a
set of effective single-particle energies
for $^{76}$Ge and  $^{76}$Se that gives, 
essentially within errors, the measured valence
orbit occupancies calculated using the SRQRPA. This effective set, at the same
time, improves  the description of the energies of low-lying $1^+$ states
in $^{76}$As.  In constructing the effective single-particle energies, we kept,
unlike in Ref. \cite{SC3}, the globally fitted
spin-orbit splittings of all orbits intact. The neutron
and proton mean field energies used previously in Refs. \cite{us} are compared
with the adjusted set used 
further here in Figs. \ref{fig:neut} and \ref{fig:prot}.

As one could see, the main difference is the overall shift of the neutron $gds$ shell closer
to the $fp$ shell. The proton levels are shifted less, with the $gds$ shell lifted
further away from the chemical potential $\lambda$ and from the $fp$ shell, i.e., an opposite tendency than
in the case of neutrons.
 The resulting occupancies of the valence subshells are shown in Table
\ref{tab:comp} and compared with the measured values \cite{Schiffer1,Schiffer2} 
(in columns 5 and 9). The occupancies
in columns 2 and 6 are those of Refs. \cite{us} evaluated from the BCS expression, Eq. (\ref{eq:BCS}),
with the standard Woods-Saxon potential as input. Those are the values quoted
in the experimental papers \cite{Schiffer1,Schiffer2}. 
The values in columns 3 and 7 are the  $n_{n(p)}^{QRPA}$ evaluated 
with the standard QRPA and using the standard Woods-Saxon potential.
We display them in order to stress the importance of proper theoretical treatment,
and to show how much difference the ground state correlations make.
Finally, in columns 4 and 8 are the occupancies that we use further here
and which  are evaluated in the correlated  ground state using the
SRQRPA and the adjusted set described above.
Note that the entries in columns 3,4,7 and 8 were evaluated with the correlated QRPA state
obtained with the coupling constant $g_{pp}$ determined in the usual way, i.e. so that the
$2\nu\beta\beta$ rate is correctly reproduced.
The sum of the corresponding entries,
i.e. the total calculated number of neutrons and protons above the $^{40}$Ca core
were already given in Table \ref{tab:ws}. 

The overall improvement in describing the occupancies is clearly visible. For neutrons, the total calculated occupancy of the valence shells is calculated to be 15.3 and 13.4
for $^{76}$Ge and $^{76}$Se, respectively, 
with 0.4(0.6) neutron vacancies in the $f_{7/2}$ and 1.0(1.2) neutrons occupying the
rest of the $gds$ shell. In particular, in the $d_{5/2}$ there should be 0.35(0.39) neutrons
according to our calculation, while Ref. \cite{Schiffer1} tentatively assigns 0.2 neutrons to $d_{5/2}$
in $^{76}$Se. 
For protons the total occupancy of valence shell is calculated to be 4.4 and 6.3, respectively, 
with 0.8(0.9) proton vacancies in $f_{7/2}$ and 0.4(0.6) protons in the rest of the $gds$ shell.

\begin{table}[htb]
\begin{center}
\caption{The calculated occupancies of individual neutron and proton orbits
for the two considered nuclei and
using BCS only in columns 2 and 6, standard QRPA in columns 3 and 7
(label Q, 
these entries were obtained 
with the standard Woods-Saxon potential) and the average nucleon number conserving
SRQRPA in columns 4 and 8 (label S, these entries were
obtained with adjusted single particle energies). In columns 5 and 9 are
the experimental occupancies of valence orbits \cite{Schiffer1,Schiffer2}. The first entry, 
for the $p$ orbit, is the sum of occupancies of $p_{1/2}$ and $p_{3/2}$ orbits.} 
\label{tab:comp}

\vspace{0.1cm}

\begin{tabular}{|c|cccc|cccc|}
\hline
 & & & $^{76}$Ge & & & & $^{76}$Se & \\
\hline
  neut. &BCS & Q & S & exp & BCS & Q & S & exp \\
 $p$ & 5.65 & 5.27 &  4.64 & 4.9$\pm$0.2 & 5.57 & 5.05 & 4.12 & 4.4$\pm$0.2  \\
 $f_{5/2}$ &  5.54 & 5.12 & 4.34 & 4.6$\pm$0.4 & 5.53 & 5.00 & 3.63 & 3.8$\pm$0.4 \\
 $f_{7/2}$ & 7.91 & 7.67 & 7.62 & - & 7.90 & 7.54 & 7.37  & - \\
 $s_{1/2}$ & 0.01 & 0.05 & 0.07  &- & 0.01 & 0.04 & 0.08 &  - \\
 $d_{3/2}$ & 0.03 & 0.14 & 0.15 & - & 0.02 & 0.14 & 0.16  &  - \\
 $d_{5/2}$ &  0.09&  0.30 &  0.36 & - &  0.07& 0.27 & 0.39  & - \\
 $g_{7/2}$ & 0.14 & 0.53 & 0.48 & - &0.12 & 0.56 & 0.58  & - \\
 $g_{9/2}$ & 4.63 & 4.78 & 6.35 & 6.5$\pm$0.3 & 2.78 & 3.55 & 5.66  & 5.8$\pm$0.3\\
 \hline
 prot. & & & & & & & & \\
 $p$ & 2.23 & 2.34 & 1.75 & 1.77$\pm$0.15 & 2.77 & 2.76 & 2.28  &  2.08$\pm$0.15 \\
 $f_{5/2}$ & 1.61 & 2.27 & 2.08 & 2.04$\pm$0.25 & 2.95 & 2.97 & 3.03 & 3.16$\pm$0.25 \\
 $f_{7/2}$ & 7.83 & 7.19 & 7.13 & - & 7.76 & 7.12 & 7.06  & -  \\
 $s_{1/2}$ & 0.00 & 0.02 & 0.03 & - & 0.00 &0.03 & 0.04 & - \\
 $d_{3/2}$ & 0.01 &0.07 & 0.07 & - & 0.01 & 0.09 & 0.09 & - \\
 $d_{5/2}$ & 0.01 & 0.12 & 0.15 & - & 0.02 & 0.17 & 0.18 & - \\
 $g_{7/2}$ & 0.02 & 0.19 & 0.16 & - & 0.03 & 0.31 & 0.27 & - \\
 $g_{9/2}$ & 0.29 & 0.85 & 0.62 & 0.23$\pm$0.25 & 0.46 & 1.15 & 1.04 & 0.84$\pm$0.25 \\
 \hline
\end{tabular}
\end{center}
\end{table}

\begin{figure}[htb]
\includegraphics[width=.48\textwidth]{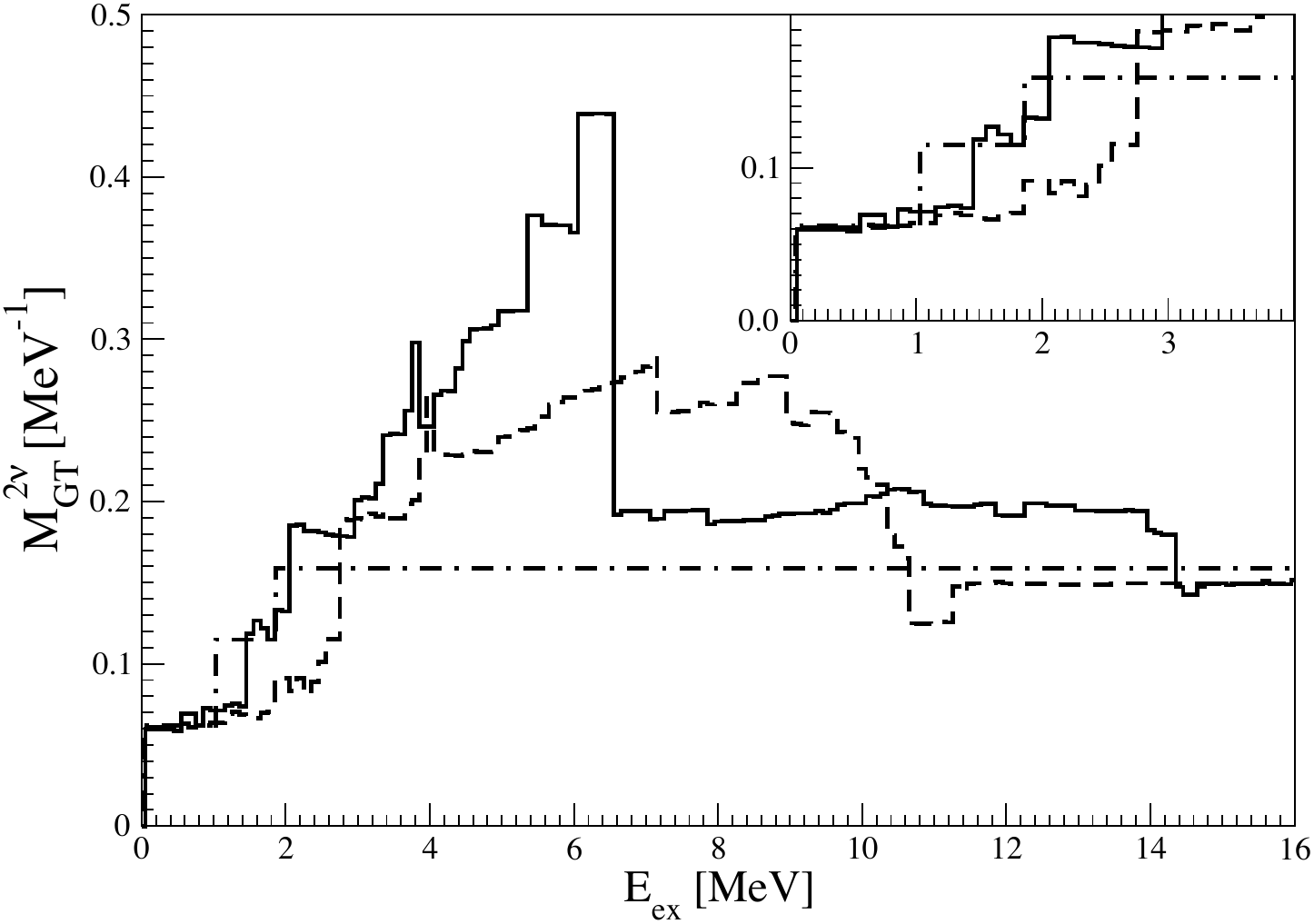}
\caption{ The running sum of $M^{2\nu}$, Eq. (\ref{eq:2nu}). The dashed line corresponds to 
Refs. \cite{us}, the full line is the present result, and the dot-and-dashed line is the experimental
result \cite{frekers}. In the insert the first 4 MeV of excitation energy are shown in detail.}
\label{fig:2nu}
\end{figure}

To further test the adequacy of the  adjusted effective mean field, the running sum of the contributions
to the $M^{2\nu}$, the nuclear matrix element for the $2\nu\beta\beta$-decay mode, is shown in
Fig. \ref{fig:2nu} and compared with the available data. The quantity displayed is
\begin{equation}
M^{2\nu} = \Sigma_{k,m, \omega_m \le \Omega} \frac{ \langle f || \vec{\sigma} \tau^+ || 1^+_k \rangle 
 \langle 1^+_k | 1^+_m \rangle \langle  1^+_m || \vec{\sigma} \tau^+ || i \rangle}
 { \omega_m - (M_i + M_f)/2} ~, 
 \label{eq:2nu}
 \end{equation}
 where on the x-axis in Fig. \ref{fig:2nu} we use the excitation energy $E_{ex}$ in the intermediate
 nucleus $^{76}$As instead of the $\Omega$ the largest included
 eigenvalue of the QRPA equation of motion.
 
We adopted the adjusted  effective mean field energies shown in Figs. \ref{fig:neut} and \ref{fig:prot}
 for $^{76}$Ge and  $^{76}$Se together with the SRQRPA method for the evaluation of the
 $2\nu\beta\beta$ nuclear matrix element. Again, the adjusted effective mean field describes
 much better the energies and amplitudes of the states for excitation energies below
 $\sim$4 MeV where experimental data are available. We should point out, however, that
 the good agreement for the product of the two weak amplitudes (the numerator of Eq.(\ref{eq:2nu}))
 does not mean that our calculation is free from the general problem of QRPA calculations,
 namely that the $\beta^-$ strength corresponding to the GT transition $^{76}$Ge $\rightarrow$
 $^{76}$As is too large while the $\beta^+$ strength $^{76}$Se $\rightarrow$ $^{76}$As is too
 small, as stressed, e.g., in Ref. \cite{Suh05}. Based on Fig. \ref{fig:2nu} one can
 conclude that the reasonable agreement between the experimental value of $M^{2\nu}$ 
 and the value based on the few low-lying states
 (the so-called low-lying states dominance) appears to be accidental, at
 least in this case; if measurements
 could be extended to $\sim$5-6 MeV and stopped there, that agreement would be lost, 
particularly when the adjusted energies are used.  
 
While the chosen procedure is not ideal, since the effective single-particle energies
are chosen ad hoc, it
is clearly an improvement when compared to our previous work \cite{us}. First, the adopted
method, SRQRPA, conserves the mean particle number in the correlated ground state,
unlike the QRPA. Second, the occupancies of the valence orbits, in both
protons and neutrons, agree now within errors with the experimental data \cite{Schiffer1,Schiffer2}.

\section{Results}

We are now in the position to ascertain to what degree the $0\nu\beta\beta$-decay nuclear
matrix element $M^{0\nu}$ for the $^{76}$Ge $\rightarrow$ $^{76}$Se transition
change with the modification of the effective
mean field energies in combination with  the application of the SRQRPA method.
In Table \ref{tab:val1} we show the sequence of the $M^{0\nu}$ values, together with
the fitted renormalization constants $g_{pp}$ as well as the calculated halflife for the
nominal neutrino mass $\langle m_{\beta\beta} \rangle$ = 50 meV. The 
Table entries were obtained using $^{40}$Ca as a core, with the $p,f$ and $s,d,g$ shells (9 single particle orbits) for both neutrons and protons included, with unrenormalized value $g_A$ = 1.25,
CD-Bonn based G-matrix interaction,
and with Jastrow-like function included \cite{MS} in order to 
include the effect of the short range correlations.
(We will discuss the dependence on these assumptions below.
Note that since we use $g_A = 1.25$ everywhere in this work, we denote the
$0\nu\beta\beta$ nuclear matrix element as $M^{0\nu}$ and do not use the
notation $M'^{0\nu} = (g_A/1.25)^2 M^{0\nu}$ of Refs. \cite{us}.)

\begin{table}[htb]
\begin{center}
\caption{The values of $g_{pp}$, $M^{0\nu}$, and $T_{1/2}^{0\nu}$ for 
 $\langle m_{\beta\beta} \rangle$ = 50 meV in units of 10$^{26}$y, for different
 variants of the calculations. For further explanations, see text.} 
\label{tab:val1}

\vspace{0.1cm}
\begin{tabular}{|l|c| c| c |}
\hline
variant &  $g_{pp}$ & $~~M^{0\nu}$~~ & $T_{1/2}^{0\nu}$(10$^{26}$y) \\
\hline
\\
\smallskip
QRPA, WS mean field \cite{us} &  ~0.849~ & ~4.34~  &  7.0 \\
\smallskip
RQRPA, WS mean field \cite{us} &0.990 & 3.81 &  9.1 \\
\smallskip
QRPA , effective mean field & 0.903 & 4.23  & 7.4 \\
\smallskip
RQRPA, effective mean field &1.170 & 3.44 & 11.1 \\
RQRPA, effective mean field  &  & & \\
\smallskip
                 ~and adjusted $M^{2\nu}$ denominators & 1.206 & 3.18 & 13.0  \\
\smallskip
SRQRPA, effective mean field & 1.125 & 3.66 & 9.8 \\
SRQRPA, effective mean field  &  & & \\
                 ~and adjusted $M^{2\nu}$ denominators & 1.184 & 3.27 & 12.3 \\
\\
\hline
\end{tabular}
\end{center}
\end{table}

The last, and thus our final entry for this variant
in Table \ref{tab:val1}, 
$M^{0\nu}$ = 3.27, was obtained when, in addition, the lowest energy denominator
in the expression for $M^{2\nu}$, Eq.(\ref{eq:2nu}), was shifted in such a way that it agreed 
with the known energy of the first $1^+$ state in $^{76}$As. All other energy denominators were
then shifted by the same amount as the first one. (The entry in line 5 was obtained this way
as well.)
This procedure is commonly used when $M^{0\nu}$ are evaluated,
but was not employed previously in Refs.\cite{us}. This  value, $M^{0\nu}$ = 3.27,
can be compared now with the
same quantity calculated by other methods. To make the comparison meaningful one has to
keep in mind that $M^{0\nu}$ contain $R$, the nuclear radius, as a factor. Unfortunately,
different authors use different conventions for $R = r_o\times A^{1/3}$. In our work
we use $r_0$ = 1.1 fm. Presumably the same is used in Ref. \cite{SC3}, 
even though the $r_0$ value is not explicitly quoted there, while in the
shell model works \cite{Poves} $r_o$ = 1.2 fm. 

With this correction included, the shell model
value \cite{Poves}  of $M^{0\nu}$ is 2.11, and the QRPA result of Ref. \cite{SC3} with the adjusted
single-particle energies chosen there, is 2.8. 
All these calculations use the same method, namely the Jastrow function
\cite{MS} for the treatment of short range correlations.
It, therefore, appears that the adjustment of mean field energies, 
in order to correctly reproduce the measured
occupancies of the valence orbits \cite{Schiffer1,Schiffer2}, results in reduction of
$M^{0\nu}$ when QRPA or its generalizations are used. The discrepancy with respect
to the shell model result \cite{Poves} is reduced, in our work,  to about half of its previous
magnitude. 

As in Refs. \cite{us} we would like to estimate the possible range of the $M^{0\nu}$ values
taking into account  changes corresponding to the variation in the number
of single particle states included and to  different treatment of the short range repulsion.
Accordingly, we repeated the $M^{0\nu}$ evaluation with 3 and 5 oscillator shells included,
in addition to the two oscillator shell result described above. The single particle energies
in these additional shells were kept at their original Woods-Saxon potential values.
For all these three variants we performed the calculations using the Jastrow fuction 
\cite{MS} for the treatment of short range correlations as well as the Unitary
Correlation Operator Method (UCOM) \cite{ucom}. We kept the axial current
coupling constant $g_A$ at its unrenormalized value 1.25 in all cases.
The previous, Refs. \cite{us}, and the new results are compared in Table \ref{tab:finalge}

 \begin{table}[htb]
\begin{center}
\caption{The calculated $M^{0\nu}$ matrix elements for the $^{76}$Ge $0\nu\beta\beta$
decay; the mean value and its range are shown for the two alternative
treatments of the short range correlations. In column 2 are the previous values
obtained with (R)QRPA method and with the Woods-Saxon potential single
particle energies \cite{us}, and in column 3 are the values obtained with the SRQRPA method
and the adjusted energies described above. }
\label{tab:finalge}

\vspace{0.2cm}

\begin{tabular}{|c|cc|}
\hline
&  prev. & new \\
\hline
Jastrow s.r.c. & 4.24(0.44) &  3.49(0.23) \\
UCOM s.r.c. &  5.19(0.54) &  4.60(0.39) \\
\hline
\end{tabular}
\end{center}
\end{table}

\section{Application to the $^{82}$ Se $0\nu\beta\beta$ decay}

It is reasonable to expect that the modifications of the mean field energies, described
above, that were
relevant to the $^{76}$Ge $\rightarrow$ $^{76}$Se decay, will also apply, at least approximately,
to the $^{82}$Se $\rightarrow$ $^{82}$Kr decay, since in that case
there are just two more protons and four more neutrons compared to the $A=76$ case.
Guided by such considerations we modified the Woods-Saxon potential energies
used previously in Refs.\cite{us} for $A=82$ for the protons  in both
 $^{82}$Se and  $^{82}$Kr as for the proton energies for $^{76}$Se, and for the neutrons
 as in the $^{76}$Ge that is closer to them in the number of neutrons.  
 The resulting valence orbit occupancies are shown in Table \ref{tab:seocc}.
 In columns 2 and 4 are the $n_{n(p)}^{QRPA}$ values evaluated with the standard
 Woods-Saxon potential and in columns 3 and 5 are the $n_{n(p)}^{SRQRPA}$ values
 evaluated with the adjusted energies.
 In the neutron system the present treatment predicts that the valence shells
 contain only $\sim$19(17) neutrons compared with the naive expectation of 20(18)
 neutrons.
 
 \begin{table}[htb]
\begin{center}
\caption{The calculated occupancies of valence neutron and proton orbits
for $^{82}$Se and $^{82}$Kr. See the text for explanation} 
\label{tab:seocc}

\vspace{0.1cm}

\begin{tabular}{|c|cc|cc|}
\hline
&    ~$^{82}$Se &   & ~$^{82}$Kr   \\
& prev. & new & prev. & new  \\
\hline
neutrons& & & &  \\
$p$ &  5.5 & 5.2 & 5.4 & 4.8  \\
$f_{5/2}$ & 5.4 & 5.2  & 5.4 & 4.9  \\
$g_{9/2}$ & 7.2 & 8.5  & 6.1 & 7.3 \\
\hline
protons & & & &  \\
$p$ & 2.7 & 2.2  & 3.2 & 2.8  \\
$f_{5/2}$ & 3.0 & 3.2  & 3.6 & 4.0 \\
$g_{9/2}$ & 1.0 & 0.9  & 1.3 & 1.2  \\
\hline
\end{tabular}
\end{center}
\end{table}

Performing the same set of calculations as described previously for the $^{76}$Ge decay,
we conclude that the $0\nu\beta\beta$ nuclear matrix element for the $^{82}$Se decay
is also reduced from the previous (R)QRPA value of 3.76(0.40) for 
Jastrow s.r.c. and 4.59(0.39) for UCOM s.r.c. to the SRQRPA values (with modified
mean field energies and shifted energy denominators for the $2\nu\beta\beta$ decay)
of  3.50(0.24) for Jastrow s.r.c. and 4.54(0.40) for UCOM s.r.c.,
again reducing somewhat the difference between this value and the shell model result
of 2.0 \cite{Poves} (adjusted for the different values of $r_0$ and with Jastrow s.r.c.).

\section{Contribution of individual orbits to $M^{0\nu}$}

In the QRPA, RQRPA, and SRQRPA the ${M}^{0\nu}$ is written as the sum
over the virtual
intermediate states, labeled by their angular momentum and parity
$J^{\pi}$ and indices $k_i$ and $k_f$:
\begin{eqnarray}
M_K  =  \sum_{J^{\pi},k_i,k_f,\mathcal{J}} \sum_{pnp'n'}
(-1)^{j_n + j_{p'} + J + {\mathcal J}} \times~~~~~~~~~~
\nonumber\\
\sqrt{ 2 {\mathcal J} + 1}
\left\{
\begin{array}{c c c}
j_p & j_n & J  \\
 j_{n'} & j_{p'} & {\mathcal J}
\end{array}
\right\}  \times~~~~~~~~~~~~~~~~~~~
\nonumber \\
\langle p(1), p'(2); {\mathcal J} \parallel \bar{f}(r_{12})
O_K \bar{f}(r_{12}) \parallel n(1), n'(2); {\mathcal J} \rangle \times~~
\nonumber \\
\langle 0_f^+ ||
[ \widetilde{c_{p'}^+ \tilde{c}_{n'}}]_J || J^{\pi} k_f \rangle
\langle  J^{\pi} k_f |  J^{\pi} k_i \rangle
 \langle  J^{\pi} k_i || [c_p^+ \tilde{c}_n]_J || 0_i^+ \rangle ~.
\nonumber\\
\label{eq:long}
\end{eqnarray}
The operators $O_K, K$ = Fermi (F), Gamow-Teller (GT), and Tensor
(T), contain neutrino potentials and spin and isospin operators, and
RPA energies $E^{k_i,k_f}_{J^\pi}$.
The ${\mathcal J}^{\pi}$ labels angular momentum and parity  of the
pairs of neutrons that are transformed into protons with the same
${\mathcal J}^{\pi}$.

The nucleon orbits are labeled in Eq.(\ref{eq:long}) by $p,p',n,n'$. We can isolate the
contribution of, say, neutron orbits $n,n'$ by fixing these two labels, but performing
the summation over all other indeces. The resulting two-dimensional array $f(n,n')$
obviously must obey $\Sigma_{n,n'}  f(n,n') =  M^{0\nu}$; the individual contributions can be
positive or negative.  
It is interesting to visualize such contributions in order to
see which orbits are important and which are not, and to gain a better understanding
of the various physics effects affecting the $M^{0\nu}$ values. 

\begin{figure}[htb]
\includegraphics[width=.48\textwidth]{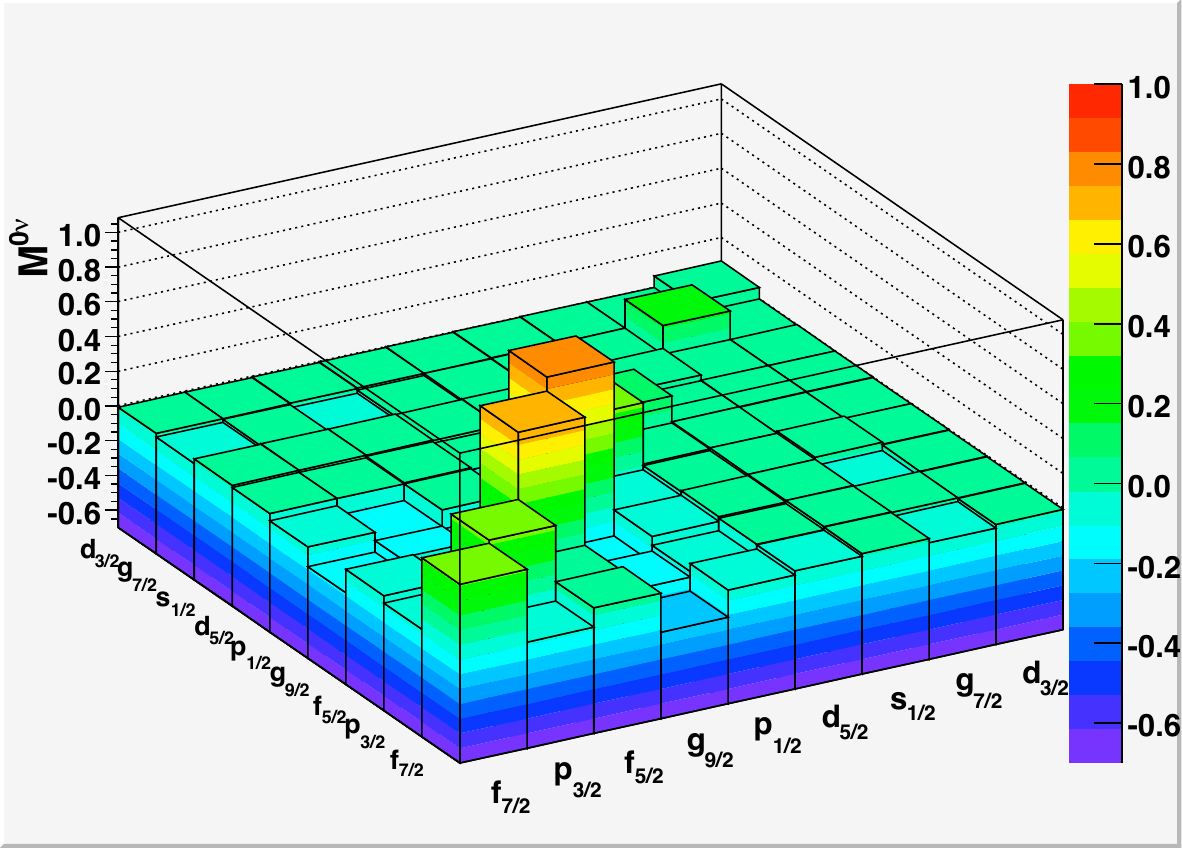}
\caption{ The contribution of individual neutron orbit pairs to $M^{0\nu}$. The entries
are normalized so that their sum is unity.}
\label{fig:legon}
\end{figure}

\begin{figure}[htb]
\includegraphics[width=.48\textwidth]{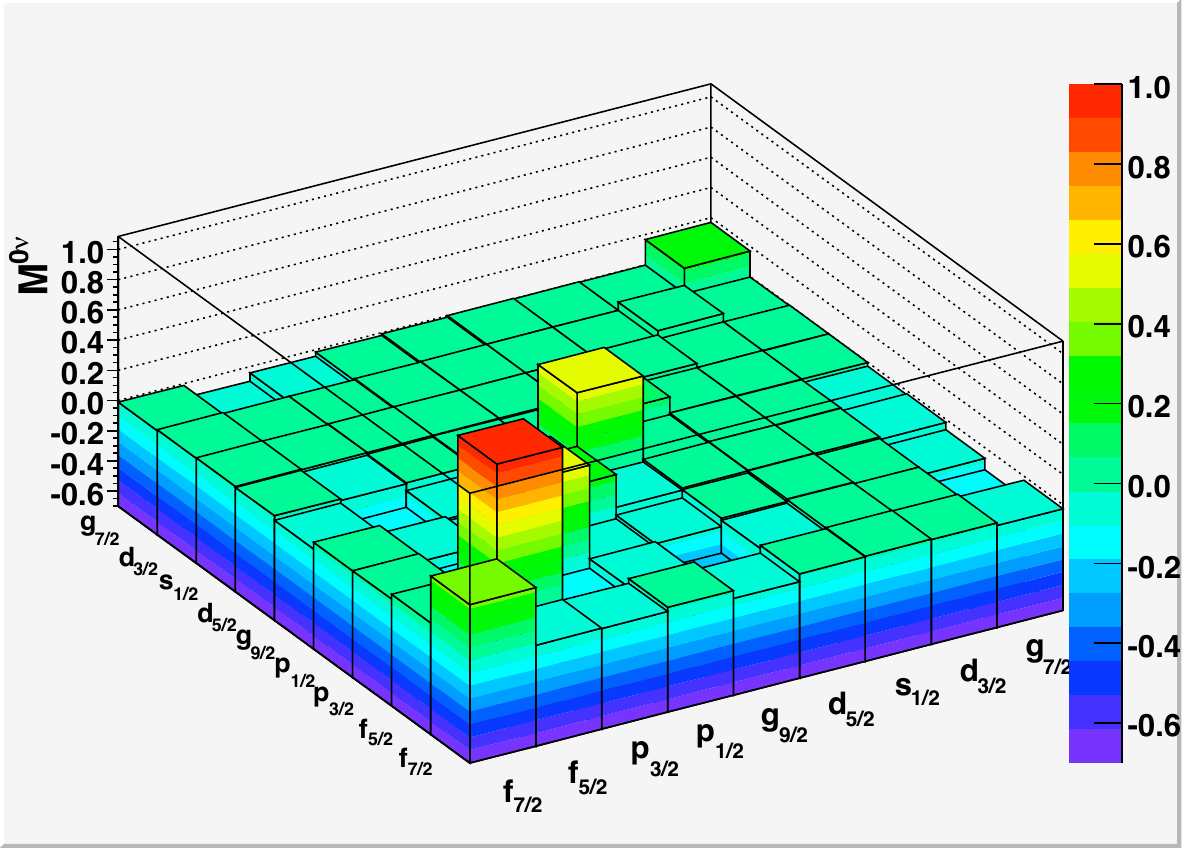}
\caption{ The contribution of individual proton orbit pairs to $M^{0\nu}$. The entries
are normalized so that their sum is unity.}
\label{fig:legop}
\end{figure}

We show a lego plot of such contributions to the $M^{0\nu}$ for the $^{76}$Ge decay, 
normalized to unity, 
in Figs. \ref{fig:legon} and \ref{fig:legop}.
The large positive
contributions along the diagonals, stemming dominantly, but not exclusively, 
from the ${\mathcal J}^{\pi} = 0^+$
pairing part of $M^{0\nu}$,
contribute  +2.97 when added together. 
The off-diagonal entries, related
to the `broken pairs' or higher seniority parts of $M^{0\nu}$, 
give  {-1.97} when added. 
The well known opposite tendencies of the pairing and broken pairs
contributions is thus clearly visible. In addition, one can also see
that the valence orbits $g_{9/2}, p_{3/2},f_{5/2},p_{1/2}$ contribute considerably
more than the orbits further away from the Fermi level, even though the $f_{7/2}$
and $g_{7/2}$ give nonnegligible contributions.

\begin{figure}[htb]
\includegraphics[width=.48\textwidth]{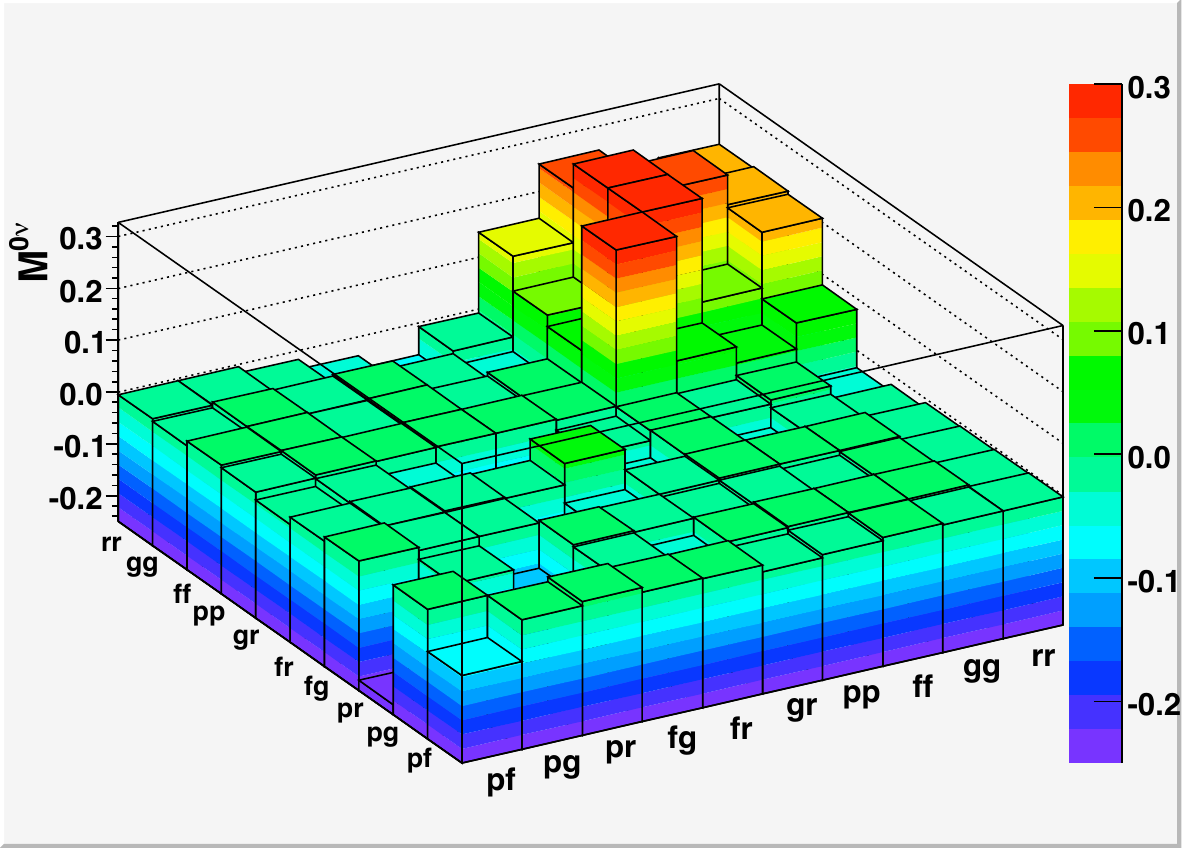}
\caption{ The contribution of  the initial neutron orbits (combination $n,n'$) (along the $y$ axis)
plotted against  the analogous combinations $p,p'$ 
(along the $x$ axis) of the final proton orbits. The entries
are again normalized to unity. For notation along the axes, see text.
}
\label{fig:legopn}
\end{figure}

In order to better visualize which  combinations of neutron and proton orbits contribute
one could, in principle, isolate in Eq.(\ref{eq:long}) the pieces corresponding to
the combination $n,n'$ of neutron orbits 
from which the neutrons disappear in the initial nucleus
and plot them against the combination $p,p'$
of proton orbits in which the protons appear in the final nucleus. 
Such a plot, however, would be difficult to visualize since it would represent a
81$\times$81 matrix even with our minimal space of nine orbits. Instead, we consider just
the three valence orbits $p$ (representing both $p_{1/2}$ and $p_{3/2}$), $f_{5/2}$ and $g_{9/2}$,
and lump all the other orbits further removed from the Fermi level into one combination
$r$ (for remote). This allows us to reduce the dimension of the matrix and the corresponding
plot to 10$\times$10, shown in Fig. \ref{fig:legopn}. Again the entries are normalized so that their
sum is unity, and the labels along the $x$ and $y$ axes are arranged in such a way that
most of the negative entries are in the front (total, naturally, again -1.97) and most of the 
positive entries are near the far corner, in order to enhance visibility.

In Fig. \ref{fig:legopn} the contribution of the $r$ non-valence 
remote orbits is sizable, and for the negative
entries, in fact, dominating. However, the positive and negative contributions from combinations
that include the $r$ orbits cancel each other  to a large extent (positive contributions total 1.29
and negative ones -1.45) so that the net effect on $M^{0\nu}$
of the remote orbits is only $\sim$15\%.

However, in QRPA and its generalizations the inclusion of orbits of at least two oscillator
shells, i.e. the set that obeys the full Ikeda sum rule, is essential. Without it, the
description of the $2\nu\beta\beta$ decay is impossible with a reasonable value of the effective
particle-particle coupling constant $g_{pp}$.

\section{Conclusion and Acknowledgments}

The occupancies of valence neutron and proton orbits, determined experimentally
in Refs.\cite{Schiffer1,Schiffer2}, represent important constraints for nuclear
models used in the evaluation of the $0\nu\beta\beta$ nuclear matrix element.
In the present work we have modified the input mean field in such a way that
the valence orbits in the model obey these constraints. Within QRPA and its
generalizations we found that it is important to also choose the variant of the basic
method that makes such comparison meaningful by conserving
the average particle number in the correlated ground state. When following this
procedure, but otherwise keeping the same steps as in our previous
evaluation of $M^{0\nu}$ within QRPA, we find that for the $^{76}$Ge $\rightarrow$ $^{76}$Se
transition the matrix element is smaller by $\sim$25\%, reducing the previously bothersome
difference with the shell model prediction noticeably. Moreover, when we assume that
analogous changes in the mean field should be applied also to the $^{82}$Se $\rightarrow$
$^{82}$Kr $0\nu\beta\beta$ decay, that differs from the $^{76}$Ge decay by only two additional
protons and four additional neutrons, we find similar reduction in $M^{0\nu}$ as well.
Clearly, having the experimental orbit occupancies available, and adjusting the input
to fulfill the corresponding constraint, makes a difference. It would be very useful to have
similar constraints available also in other systems, in particular for $^{130}$Te and/or $^{136}$Xe.

\medskip

We would like to thank Professor John Schiffer for numerous enlightening discussions.
P.V. thanks the Institute for Nuclear Theory at the University of Washington for its hospitality
during the work on this paper.  We acknowledge also the support of  the EU ILIAS project 
under the contract RII3-CT-2004-506222, the Deutsche Forschungsgemeinschaft (436 SLK 17/298)
and of the VEGA Grant agency of the Slovak Republic under the contract 
No.~1/0249/03.


\newpage

\begin{thebibliography}{99}

\bibitem{APS} S. J. Freedman and B. Kayser, {\it The Neutrino Matrix},  physics/0411216.

\bibitem{AEE} F. T. Avignone, S. R. Elliott and J. Engel, Rev. Mod. Phys. {\bf 80}, 481 (2008).

\bibitem{us} V. A. Rodin, A. Faessler, F. \v{S}imkovic and P. Vogel,
Phys. Rev. C{\bf 68}, 044302(2003); V. A. Rodin, A. Faessler, F. \v{S}imkovic and P. Vogel,
Nucl. Phys. {\bf A766}, 107 (2006), and erratum {\bf A793}, 213 (2007); 
F. \v{S}imkovic, A. Faessler, V. A. Rodin, P. Vogel, and J. Engel Phys. Rev. C{\bf 77}, 045503(2008).

\bibitem{SC}J. Suhonen, Phys. Lett. {\bf B607}, 87 (2005); 
J. Suhonen and O. Civitarese, Phys. Lett. {\bf B626}, 80 (2005);
ibid Nucl. Phys. {\bf A761}, 313 (2005).

\bibitem{Schiffer1} J.P.Schiffer {\it et al.}, Phys. Rev. Lett. {\bf 100}, 112501 (2008).

\bibitem{Schiffer2} B. P. Kay {\it et al.}, arXiv:0810.4108.

\bibitem{BM} A. Bohr and B. Mottelson {\it Nuclear Structure, Vol. I}, V. J. Benjamin, New York, 1969.

\bibitem{Bertsch} G. F. Bertsch {\it The Practitioners Shell Model}, North Holland, New York, 1972.

\bibitem{SC2} M. Aunola and J. Suhonen, Nucl. Phys. {\bf A602}, 133 (1996).

\bibitem{SC3} J. Suhonen and O. Civitarese, Phys. Lett. {\bf B668}, 277 (2008).

\bibitem{Poves} J. Men\'{e}ndez, A. Poves, E. Caurier, and F. Nowacki, arXiv: 0801.3769(nucl-th);
E. Caurier, F. Nowacki and A. Poves, Eur. Phys. J. {\bf E17},1(2008).

\bibitem{Delion} D. S. Delion, J. Dukelsky and P. Schuck, Phys. Rev. C{\bf 55}, 2340 (1997);
F. Krmpotic {\it et al.}, Nucl. Phys. {\bf A637}, 295 (1998).

\bibitem{Schuck} J. Dukelsky and P. Schuck, Phys. Lett. {\bf B387}, 233 (1996).

\bibitem{Bobyk1} A. Bobyk, W. A. Kami\'{n}ski and P. Zar\c{e}ba, Eur. Phys. J. {\bf A5}, 385 (1999).

\bibitem{Bobyk2} A. Bobyk, W. A. Kami\'{n}ski and F. \v{S}imkovic, Phys. Rev. C{\bf 63}, 051301(R)  (2001).

\bibitem{Benes} P. Bene\v{s}, F. \v{S}imkovic, A. Faessler and W. A. Kami\'{n}ski,
Progr. Part. Nucl. Phys. {\bf 57}, 257(2006).

\bibitem{frekers} E. W. Grewe and D. Frekers, Progr. Part. Nucl. Phys. {\bf 57}, 260(2006);
D. Frekers, private communication.
 
\bibitem{Suh05} J. Suhonen, Phys. Lett. {\bf B607}, 87 (2005).

\bibitem{MS} G. A. Miller and J. E. Spencer, Ann. Phys. {\bf 100}, 562 (1976).

\bibitem{ucom} H. Feldmeier, T. Neff, R. Roth, and J. Schnack,
Nucl. Phys. {\bf A632}, 61 (1998).

\end{thebibliography}
\end{document}